\def\preprint{0}                
\def\preprint{1}                
\def\comment#1{}
\if\preprint1
        \documentclass[usenatbib]{mn2e}
        \usepackage{times}
        \usepackage{graphicx}
        \usepackage{calc}

\else
        \documentstyle[astrop-bib,referee,times]{mn}
        \newcommand{\includegraphics}[1]{}
\fi

\def\oversim#1#2{\lower0.5pt\vbox{\baselineskip0pt \lineskip-0.5pt
     \ialign{$\mathsurround0pt #1\hfil##\hfil$\crcr#2\crcr\sim\crcr}}}
\def\gsim{\mathrel{\mathpalette\oversim>}}    

\title[BH~Cru]
{Period and chemical evolution of SC stars}
\author[Albert~A.~Zijlstra et al.]
       {Albert~A.~Zijlstra$^1$\thanks{E-mail: \tt a.zijlstra@umist.ac.uk},
        Timothy~R.~Bedding$^2$\thanks{E-mail: \tt bedding@physics.usyd.edu.au},
       Andrew~J.~Markwick$^1,3$,
    Rita Loidl-Gautschy$^4$, \newauthor    Vello Tabur$^2$, 
    Kristen D. Alexander$^2$, Andrew P. Jacob$^2$, 
    L\'aszl\'o~L. Kiss$^2$,    Aaron Price$^5$,  
  \newauthor Mikako Matsuura$^1$ and
           Janet~A. Mattei$^6$\thanks{Janet Mattei died on 22 March, 2004.
    The other authors would like to dedicate this paper to her memory}\\
        $^1$UMIST, Department of Physics, P.O. Box 88, Manchester M60 1QD, UK\\
        $^2$School of Physics, University of Sydney 2006, Australia\\
        $^3$Space Science Division, NASA Ames Research Center, MS 245-3, 
           Moffett Field, CA 94035, USA  \\
        $^4$Basel, Switzerland  \\
        $^5$AAVSO, 25 Birch St., Cambridge, MA 02138, USA \\
}
\pubyear{2004}

\begin{document}

\maketitle

\begin{abstract}
  The SC and CS stars are thermal-pulsing AGB stars with C/O ratio close to
  unity. Within this small group, the Mira variable BH Cru recently evolved
  from spectral type SC (showing ZrO bands) to CS (showing weak C$_2$).
  Wavelet analysis shows that the spectral evolution was accompanied by a
  dramatic period increase, from 420 to 540 days, indicating an expanding
  radius.  The pulsation amplitude also increased.  Old photographic plates
  are used to establish that the period before 1940 was around 490 days.
  Chemical models indicate that the spectral changes were caused by a decrease
  in stellar temperature, related to the increasing radius. There is no
  evidence for a change in C/O ratio. The evolution in BH Cru is unlikely to
  be related to an on-going thermal pulse.  Periods of the other SC and CS
  stars, including nine new periods, are determined. A second SC star, LX Cyg,
  also shows evidence for a large increase in period, and one further star
  shows a period inconsistent with a previous determination.  Mira periods may
  be intrinsically unstable for C/O$\,\approx 1$; possibly because of a
  feedback between the molecular opacities, pulsation amplitude, and period.
  LRS spectra of 6 SC stars suggest a feature at $\lambda>15\mu$m, which
  resembles one recently attributed to the iron-sulfide troilite. Chemical
  models predict a large abundance of FeS in SC stars, in agreement with the
  proposed association.

\end{abstract}


\section{Introduction}

The observed properties of stars on the Asymptotic Giant Branch (AGB) are
largely determined by the ratio of carbon to oxygen in their atmospheres.
Carbon abundances in AGB stars increase due to dredge-up following their
regular thermal pulses.  After a series of these events, which happen every
$10^4$--$10^5$\,yr, the enhanced carbon may cause a transition from an
oxygen-rich star to a carbon-rich star, via the intermediate S-stars
\citep[e.g.,][]{LE84}.

The SC stars \citep{KB80} form a continuous spectral sequence intermediate
between the S and C stars \citep{CF71}.  They show very strong sodium D-lines,
and strong CN bands \citep{S73}, and either weak ZrO bands (SC stars) or C$_2$
bands (CS stars). The molecular abundances indicate a C/O number ratio very
close to unity, so that CO formation leaves little C or O for the other
molecules.

The SC star BH~Crucis has simultaneously been found to show a unique
combination of a lengthening period \citep*{BMV88, WM91, WIW95}, and evolution
in spectral type from SC to CS \citep{LE85}, with ZrO bands \citep{CF71}
disappearing and C$_2$ band appearing.  The spectral evolution has been
interpreted as due to an increase in C/O ratio \citep{Whi99}.  Both the period
and abundance changes have been suggested to be caused by a recent thermal
pulse \citep{Whi99,WZ81}. BH Cru could therefore present a unique case of
real-time AGB evolution.

\begin{figure*}
\includegraphics[width=0.8\textwidth, clip=true]{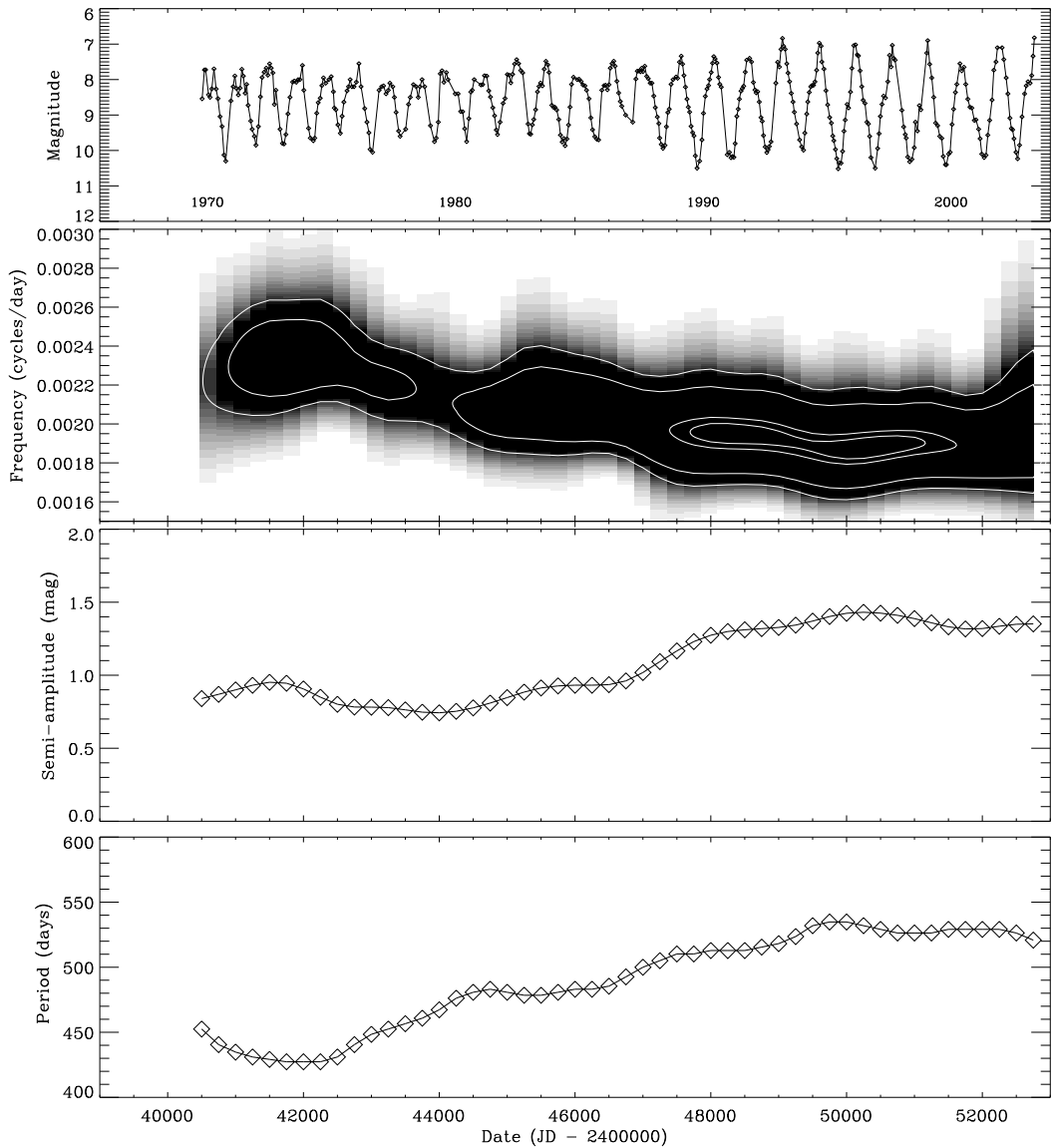}
\caption{\label{bhcru.ps} The wavelet analysis for BH~Cru, 1970--2002,
based on visual observations. Shown
are: the light curve, the frequency, the semi-amplitude of the main
frequency component and its period.}
\end{figure*}

In this paper we present an analysis of the light curve of BH~Cru, to quantify
the period and amplitude variation.  Chemical modelling is used to investigate
whether the change in relative molecular abundances can be explained as due to
a decrease in effective temperature, at constant C/O ratio.  We also present
new periods for other SC stars, including two stars which may show period
changes similar to BH Cru.

\section{Pulsation in BH~Cru}

\subsection{The star}

BH~Cru (also known as He 120) was reported to be variable by \citet{Welch69},
and assigned a spectral type of SC\,4.5-7 \citep{K71, KB80}.  The star is
bright, reaching a visual magnitude of 7, with an amplitude between 1$^{\rm
  m}$ and 3$^{\rm m}$.  \citet{WMF00} suggest a distance of $d=900\,\rm pc$,
from the Mira $PL$ relation.  At Galactic metallicity, carbon stars form for
initial masses $\gsim 2\,\rm M_\odot$ \citep[][their Table 5]{GdJ93}.  The
galactic $z$-height of BH~Cru of about 100\,pc is consistent with such a
progenitor star. A strong Sr\,I 4607\AA\ line shows evidence for s-process
element enhancements.

Optical photometry from \citet{Walker79} and infrared photometry from
\citet{WMF00} are listed in Table \ref{ir.dat}. The colours are typical for
unobscured carbon stars, and the IRAS flux densities are consistent with
photospheric emission.  The K$-$[12]$=0.96$ colour gives a black-body
temperature of 3000\,K, identical to the stellar temperature (section
\ref{model} below).  There is no evidence for on-going mass loss, neither from
self-obscuration in the near-infrared, nor from a mid-infrared dust excess.
The lack of mass loss is unusual for such a long-period Mira.

\begin{table}
\caption[]{\label{ir.dat} Visual and infrared magnitudes of BH~Cru, as
given in \citet{Walker79} and \citet{WMF00}. 
  }
\begin{flushleft}
\begin{tabular}{llllllllllllllll}  
\hline
Filter & U     & B     &  V   &  R   &  I  & 
 J   &  H     \\
  &  K  &  L  & $A_V$ 
  & $F_{12}$  & $F_{25}$  & $F_{60}$   \\
\hline
BH~Cru & 13.03 & 10.12 & 7.41 & 5.94 & 3.32  
 & 3.24 &  2.07 \\
  &  1.58 &   1.25  & 0.5
 & 20.9\,Jy & 6.6\,Jy  & 1.6\,Jy \\
\hline \\
\end{tabular}
\end{flushleft}
\end{table}

\subsection{Recent period evolution}

We have analysed visual observations collected by the Royal
Astronomical Society of New Zealand (RASNZ).  Only data from individual
observers contributing 15--20 observations or more for each star were used.
The light curve was binned into 20-day averages and analysed using wavelet
transforms, which have the advantage of being sensitive to changes of the
pulsation properties (period, amplitude) over time (e.g., \citealt{BZJF98}).
We have used the weighted wavelet Z-transform (WWZ) \citep{Fos96} developed at
the AAVSO specifically for unevenly sampled data.

The visual light curve is shown in the top panel of Fig. \ref{bhcru.ps}.
During the 1970s, the maxima tended to be flat and long-lived, with pronounced
but shorter-lasting minima. From 1990 onward, the maxima became more
pronounced and slightly brighter, with a hump on the ascending part that is
very common among Mira variables.  In the first few years, the light curve
also showed a secondary minimum superposed on the plateau of maximum. Double
maxima and minima of alternating depth are  not uncommon
among the longer-period Miras, e.g. R~Cen \citep*{HMF01}.  In R Dor
\citep{BZJF98},  a split maximum preceded a change in pulsation mode.

The wavelet plot of BH~Cru confirms the period evolution first reported by
\citet{BMV88}.  The second panel of Fig.~\ref{bhcru.ps} shows the pulsation
frequency as function of time, where the contours represent the significance
level.  The `best' period is plotted in the bottom panel.  {}From 1970 until
around 1995, we see a steady increase in the period. Since 1995, the period
has been more or less constant.  The period was 425 days at the time of
discovery (1969),  500 days in 1990 and 535 days in
1999. This large increase of over 25\%\ within (at most) 20 years is
unparallelled: among known Miras, only T UMi \citep{GS95a,MF95} has shown a
comparable, but opposite, change. The new case of LX Cyg \citep{TMP2003} is
discussed further below.

Although the period evolution shows some possible fluctuations, it can be well
fitted with a constant rate of change of 5 days per year, corresponding to
1.4\%\ per cycle, between 1975 and 1995. (The pre-1970 part of the curve shows
a reverse change, but this may be affected by edge effects in the wavelet
analysis.)

The semi-amplitude (third panel of Fig \ref{bhcru.ps}) shows similar
strong evolution.  It was around 0.8 mag before 1980 and, in parallel
with the period increase, the semi-amplitude rapidly increased to 1.4
mag in about 1992.  Except for a single faint maximum in AD 2000, the
amplitude has since remained constant.

\begin{figure}
\includegraphics[width=8.45cm,clip=true]{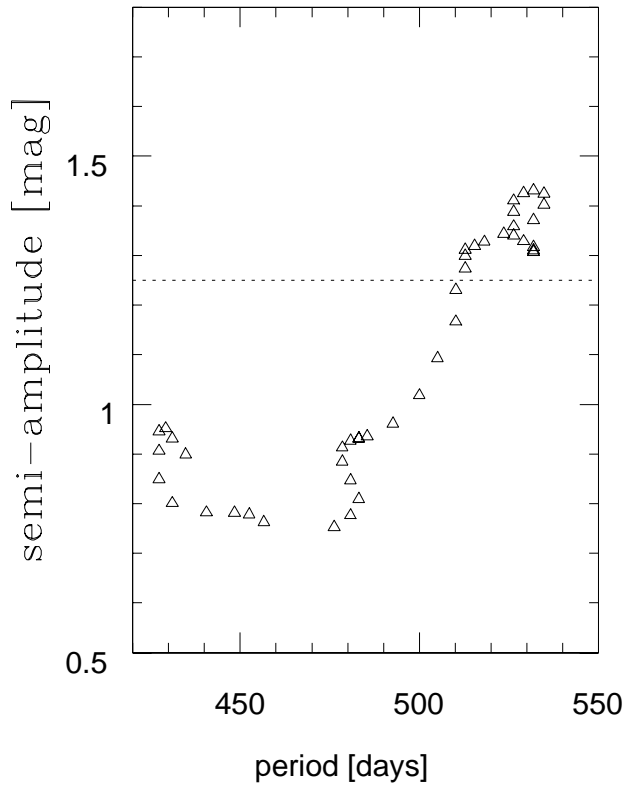}
\caption{\label{amp.ps} The visual semi-amplitude, in magnitudes, versus the 
period in days, since JD 2441000 (AD 1970). The dotted line shows the
lower limit of the Mira classification. }
\end{figure}

Fig. \ref{amp.ps} illustrates the correlation between the increases in
amplitude and period. It shows a roughly linear correlation, with the
semi-amplitude increasing by about 50 mmag/cycle, or about 5\%\ per
cycle. The correlation between period and semi-amplitude evolution mirrors
that seen in R Hya \citep*{ZBM02}, and also in R~Aql and S~Ori
\citep{BCZ2000} and Y~Per \citep{KSS2000}.

\subsection{Older period data}

Although BH Cru was only discovered to be variable in 1969, the region has
been extensively covered by older surveys. But we have not found published
records of BH~Cru prior to 1969. The star is absent from the CP catalogue
\citep{GK1900} but this has a completeness limit of 9.2\,mag and the blue
magnitude of BH~Cru is always below this. The deeper CD catalogue
\citep{Tho1932} goes to 10\,mag, but also does not include it.  These limits
do not allow for an unambiguous interpretation. We have therefore searched the
Harvard plates for evidence of its pre-discovery behaviour.

\begin{figure}
\includegraphics[width=8.45cm, clip=true]{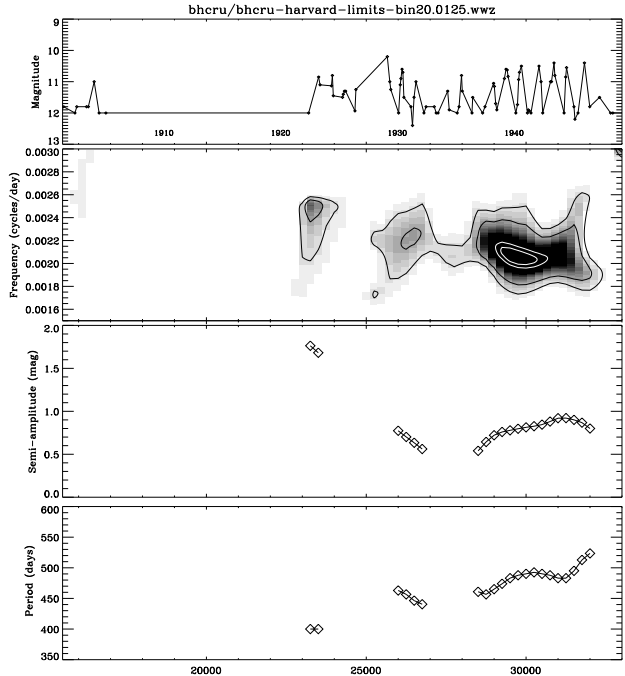}
\caption{\label{oldp.ps} Photographic observations of BH Cru in 20-day bins.
Upper limits, where a comparison star of the that magnitude could be
seen, are included if at m=11.8, but brighter limits are not. The
systematic photometric accuracy of the data is no better than 0.5
mag. Period and amplitude before JD2425000 have very low confidence.}
\end{figure}

The Harvard Plate Stack contain 96 good observations of BH Cru, 15
detections with uncertain magnitude, and 37 plates where the star is not
visible.  The plates are almost exclusively blue. Combined with BH Cru's
$B-V$, the star exists near the limit of the plates while at maximum. The
data are shown in Fig. \ref{oldp.ps}. We performed a wavelet analysis using
the 96 good and 15 uncertain determinations, and also included upper
limits where they were fainter than 11.8mag. (The brighter limits do not
constrain the variability.)

BH Cru was a pulsating variable well before its discovery as such.  Strong
pulsations are evident, with an amplitude between 1 and 2 magnitude. The
plates before 1910 are  consistent with little or no fluctuations, but
the sampling is poor and this is not conclusive. The data are too patchy for a
well-determined period, however a period of about 490 days is indicated around
1940. The period at 1930 may have been slightly shorter (about 440 days).

Compared to Fig. \ref{bhcru.ps}, the period around 1940 was about 10 per cent
longer than in the early 1970's. The data suggest that the period was not
constant even in this early phase and there may have been a further increase
shortly after 1940. Further data would be desirable, but the conclusion from
the available data is that between 1945 and 1970 the period must have
decreased.  The recent period instability may be part of a recurrent
behaviour.

\subsection{Physical changes in BH~Cru}
\label{phys}

The pulsation equation for Mira variables is
\begin{equation}
\label{over}
 \log P = 1.5\log R -0.5 \log M + \log Q,
\end{equation}
\noindent
for first overtone pulsators, where the pulsation constant $Q \approx 0.04$ 
\citep{FW82},  or
\begin{equation}
 \log P = 1.949 \log R -0.9 \log M - 2.07
\end{equation}
\noindent for fundamental mode pulsators \citep{Woo90}. $P$ is the
period in days and $R$ and $M$ are the radius and mass in solar units. 

In Mira variables, $M$ decreases on times of scales of $10^5$--$10^7$\,yr
because of the stellar wind, but over a century it is approximately constant.
In any case, BH~Cru shows no evidence for large mass loss. $Q$ can be assumed
as constant on this short time scale.  The radius $R$ is not directly known;
it is however the only parameter whose changes can explain the period
evolution of BH~Cru.

Miras are thought to be fundamental-mode pulsators, although this is not as
well determined for the very long period variables.  However, the observed
optical radii for Miras are much larger than those predicted from the
fundamental mode, and agree better with first overtone pulsators
\citep*{HST95}.  The discrepancy may be due to missing opacity in the
atmosphere models.  \citet{vBTC02} find increased diameters also at K, which
they attribute to circumstellar (water) emission. Here we prefer Eq.
(\ref{over}) since it better fits the atmospheres.

The period of BH Cru increased by 25\%\ since 1970, corresponding to a
16\%\ change in the radius from Eq. (\ref{over}).  For a stable bolometric
magnitude, this would decrease the effective temperature by 7.5\%: $\sim
250\,$K at the temperature of 3200\,K \citep*{LLJ01}.  Fundamental-mode
models require a smaller radius change (12\%) and temperature change ($\sim
200\,$K).

If the luminosity also increased, a smaller temperature decrease would be
required.  The mean magnitude of the infrared light curve changed by no more
than 0.1\,mag, between 1980 and 1995, and the J$-$K colour reddened by no more
than 0.05 mag. The bolometric correction at K therefore decreased by less than
0.05\,mag, using Fig. 16 of \citet{WMF00}.  The mean magnitude of the optical
light curve was constant to within 0.5\,mag but it is much more sensitive to
temperature variations and varying molecular opacity.  Overall, the data is
consistent with evolution at constant luminosity; any increase in the
luminosity did not exceed 10\%.

Assuming an initial radius of 300\,R$_\odot$, BH~Cru will have
expanded by $6 \times 10^{10}\,\rm m$, at an average velocity of $\sim
100\,\rm m\,s^{-1}$. Typical small-amplitude Mira variables expand at
$3$--$10\,\rm km\,s^{-1}$ during the pulsation cycle. The small
average expansion velocity would therefore not be expected to
significantly disrupt the pulsational behaviour.

\section{Equilibrium chemistry: Z\lowercase{r}O and C$_2$}
\label{chemistry}

\subsection{Spectral  changes}

The spectral evolution of BH~Cru is remarkable.  Before 1973, BH Cru showed
evidence for ZrO, but lacked C$_2$ lines \citep{CF71}; the ZrO line was weak
or absent around 1970 \citep{K71} but strong in 1967 \citep{S73}; \citet{CF71}
reported it as weakly present on an undated plate.  In 1980, lines of C$_2$
had appeared and the ZrO lines were no longer present \citep{LE85}. The
spectra taken around 1980 cover a substantial part of the light curve:
cyclical variability cannot account for the observed evolution \citep{LE85}.
The few spectra taken around 1970 show ZrO lines of variable strength: during
some part of the pulsation cycle, the pre-1973 spectra may have been lacking
in ZrO as well. But the strong ZrO bands visible at times before 1973 did not
re-occur over any part of te pulsation cycle in 1980.

The change has been interpreted as tracing an increase in the C/O ratio from
$<1$ to $>1$ \citep{Whi99}, associated with the dredge-up of carbon produced
in a helium shell flash (a thermal pulse). However, below we investigate
whether the change in effective temperature of the star can also explain the
abundance changes, without requiring a change in the C/O ratio.  We note that
a possible recurrent shift in spectrum between the presence of ZrO and C$_2$
has been seen in TT Cen \citep{S73}.

A temperature effect is indicated by the fact that SC stars tend to have
earlier subtypes than CS stars.  Only two SC stars (see Appendix \ref{appA})
are known to reach types later than SC5 (BH Cru and LX Cyg), while three of
the five CS stars are classed as C7 or later.  The temperature zero point for
C and S stars subclasses differ: C5 corresponds in temperature to about M0.
However, SC7 and C7 have similar temperatures \citep{KB80}, and the late CS
stars have therefore lower stellar temperatures than any of the SC stars with
known subclass.

\subsection{Temperature dependence}

\begin{figure}
\includegraphics[width=8.45cm,clip=true]{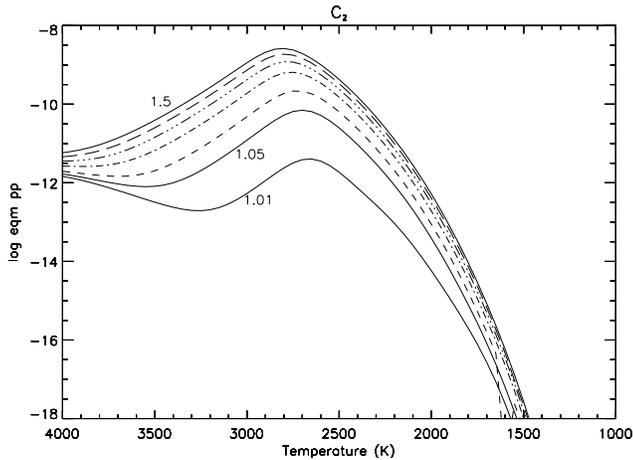}
\caption{\label{c2.ps} The equilibrium partial pressure of C$_2$ (log), 
as function of temperature, for seven different values of the C/O ratio: 
1.01, 1.05, 1.1, 1.2, 1.3, 1.4 and 1.5 (lower curve to upper curve 
respectively)}
\end{figure}

\begin{figure}
\includegraphics[width=8.45cm,clip=true]{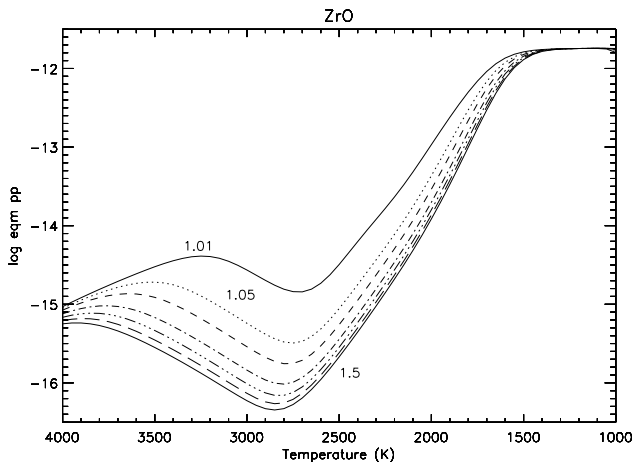}
\caption{\label{zro.ps}  Same as Fig.~\ref{c2.ps}, but for ZrO }
\end{figure}

To investigate the temperature and C/O dependence in isolation of other 
parameters, we
model the photospheric composition as a molecular equilibrium
\citep[e.g.,][]{Tsuji73, WC78, Tarafdar87, SH90}.  We calculate the molecular
equilibrium composition by a direct minimisation of the Gibbs free energy of
the system using a steepest descent technique \citep*{WJD58}.  The free energy
of the system is given by
\begin{equation}
 G=\sum_{i} f_{i}x_{i},
\end{equation}
\noindent where $x_{i}$ is the number of moles of molecule $i$ and $f_{i}$ is 
the chemical potential of molecule $i$,
\begin{equation}
f_{i}=\left(\frac{G}{RT}\right)_{i} + \ln P 
  + \ln\left(\frac{x_{i}}{\bar{x}}\right).
\end{equation}
\noindent Here, $\left({G}/{RT}\right)_{i}$ is the Gibbs free energy of 
formation of molecule~$i$, $P$~is the total pressure of the system (a
model parameter), and $\bar{x}=\sum_{i}x_{i}$. The Gibbs free energy
of formation of a molecule is a temperature-dependent thermo-chemical
property that can be found  for example in the JANAF
tables \citep{Chase85}. To estimate the value at other temperatures,
we used an interpolation formula. \citet{SH90} provide coefficients to
a polynomial fit of the form
\begin{equation}
 \Delta G = aT^{-1}+b+cT+dT^{2}+eT^{3},
\end{equation}
\noindent to the Gibbs free energies for 210 species using JANAF data.  
We added further species appropriate to AGB stars in the same way.

At chemical equilibrium, $G$ is a minimum \citep{Zemansky57}.  The equilibrium
composition is therefore the set of mole numbers $X=\left\{
  x_{i}\right\}_{i=1,n}$ which minimises $G$.  Full details of the
minimisation technique subject to these constraints can be found in
\citet{Markwick2000}.

\begin{figure}
\includegraphics[width=8.45cm,clip=true]{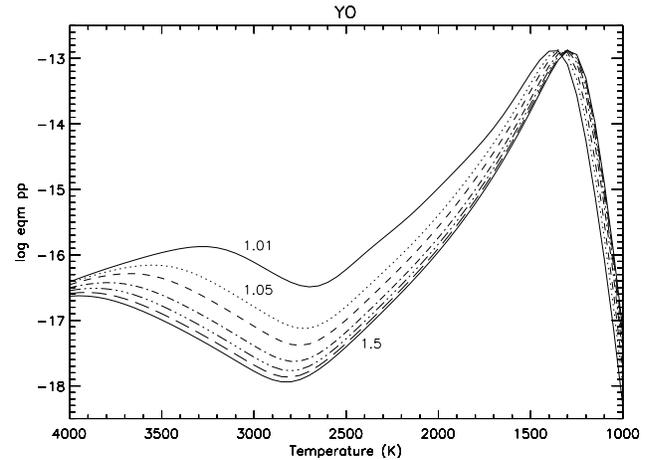}
\caption{\label{yo.ps}  Same as Fig.~\ref{c2.ps}, but for YO }
\end{figure}

\begin{figure}
\includegraphics[width=8.45cm,clip=true]{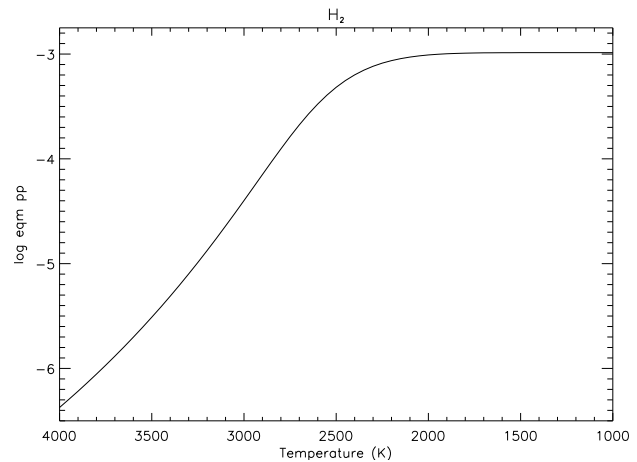}
\caption{\label{h2.ps} Same as Fig.~\ref{c2.ps}, but for H$_2$.  Molecular
hydrogen is so dominant in the model that the variation with C/O ratio is
negligible. }
\end{figure}

We adopt solar elemental abundances \citep{AG89,Cameron73} except for carbon
which is enhanced to vary the C/O ratio.  Models were run for  values
of the C/O ratio, between 1.01 and 1.5.  The resulting equilibrium abundances
of the molecules C$_2$, ZrO and YO for a variety of temperatures and C/O
ratios are shown in Figures \ref{c2.ps}--\ref{yo.ps}. A plot of the H$_2$
partial pressure is given in Fig.~\ref{h2.ps} to allow conversion of the
partial pressures into fractional abundances.

Figure~\ref{c2.ps} shows clearly that the photospheric abundance of C$_2$
increases with decreasing temperature over the range
4000--2800\,K. Furthermore, the model shows that oxides such as ZrO
can still exist at C/O ratio $>1$, although at reduced abundance
(Fig. \ref{zro.ps}). The relative abundances shown in the figure do
not take the s-process enhancements into account. In BH~Cru,
[Zr/H]$\,=1.2$ \citep{AW98}: assuming a linear scaling, the ZrO and
C$_2$ relative abundances are almost equal for C/O$\,=1.01$ at
$T=3300\,\rm K$.  \citet{AW98} give C/O$\,=1.02$.

The photospheric temperature of BH~Cru is taken as 3200\,K
(\citealt{LLJ01}; these authors found a temperature close to
minimum of about 2800\,K).  At these temperatures, Fig.~\ref{c2.ps}
shows that even a relatively small decrease in temperature leads to a
large increase in the abundance of C$_2$: a decrease of 300\,K can
yield a ten-fold increase. The peak abundance is reached at $T \approx
2800\,$K.

The disappearance of ZrO and appearance of C$_2$ took place between 1973 and
1980 \citet{LE85}.  Assuming a constant rate of period evolution, the period
increased by $\sim 50\,$days or roughly 12\% over this time.  This corresponds
to a radius increase of 8\% and a temperature decrease of 4\%\ if the
luminosity remained constant (assuming first overtone pulsation). The $\Delta
T= 130\,\rm K$ for a current temperature of 3200\,K would have increased the
C$_2$ abundance by a factor of 3--5 and decreased the ZrO abundance by a
similar factor.  The YO abundance should have decreased as well (see
Fig.~\ref{yo.ps}), but this molecule has only been weakly detected
\citep{KB80} and there is no observational mention of a change with time.

\subsection{Atmosphere models}
\label{model}

The model above has two shortcomings; we do not consider the extended
atmosphere, and no time-dependent chemistry is included.  
Chemical equilibrium is unlikely to hold at the conditions and
dynamic time scales within an extended pulsation-driven atmosphere.
However, we do not attempt to address time-dependent chemistry here.

\begin{figure*}
\includegraphics[width=17cm,clip=true]{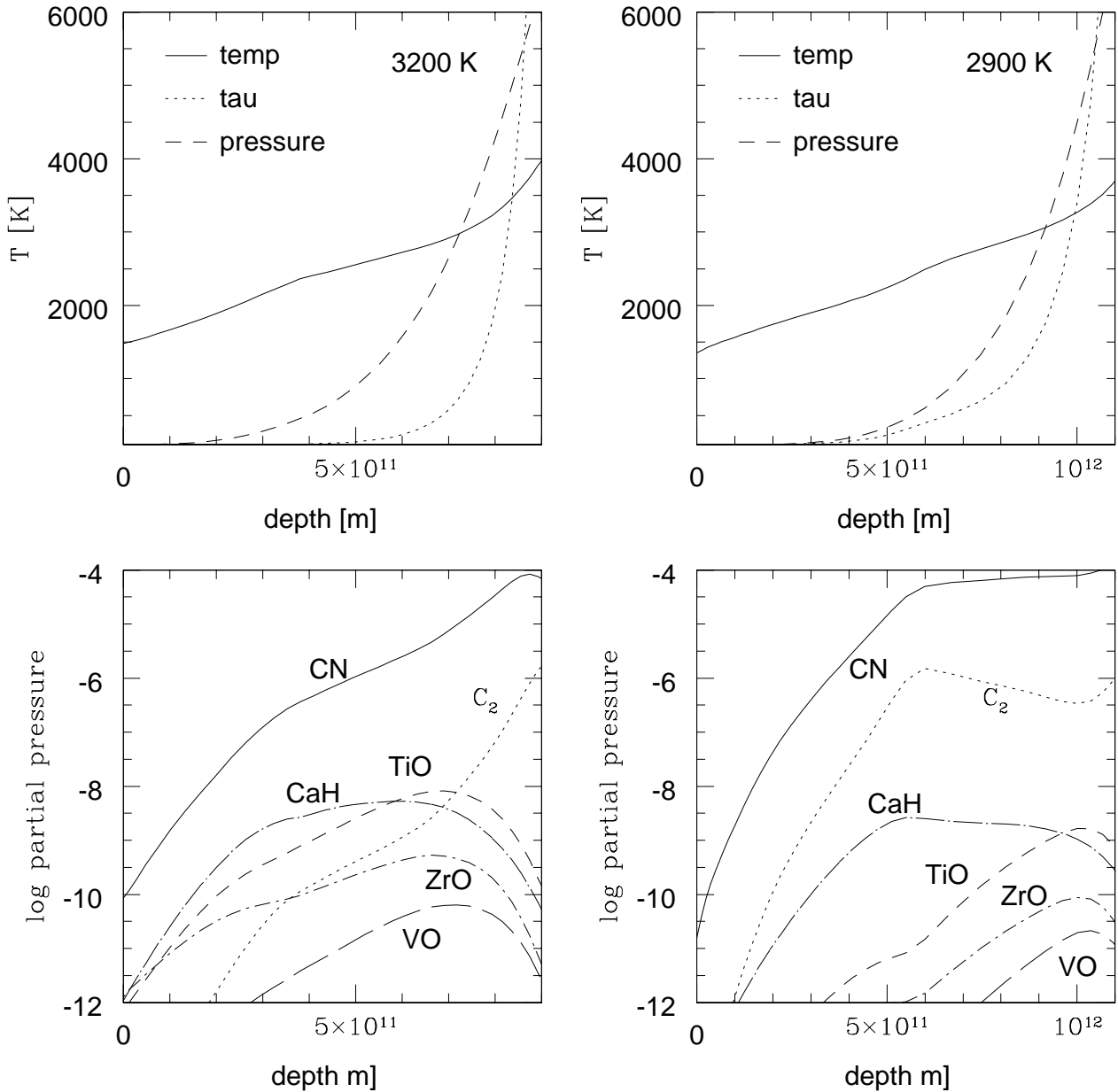}
\caption{\label{atmo.ps} Results of the atmosphere models. Top-left:
  temperature, density and opacity for a stellar temperature of 3200\,K. The
  stellar surface is on the right. The opacity scale runs from 0 to 1. 
Top-right: same, for a stellar temperature
  of 2900\,K. Bottom: Molecular partial pressures for the two temperatures. }
\end{figure*}

\begin{figure*}
\includegraphics[width=\textwidth,clip=true]{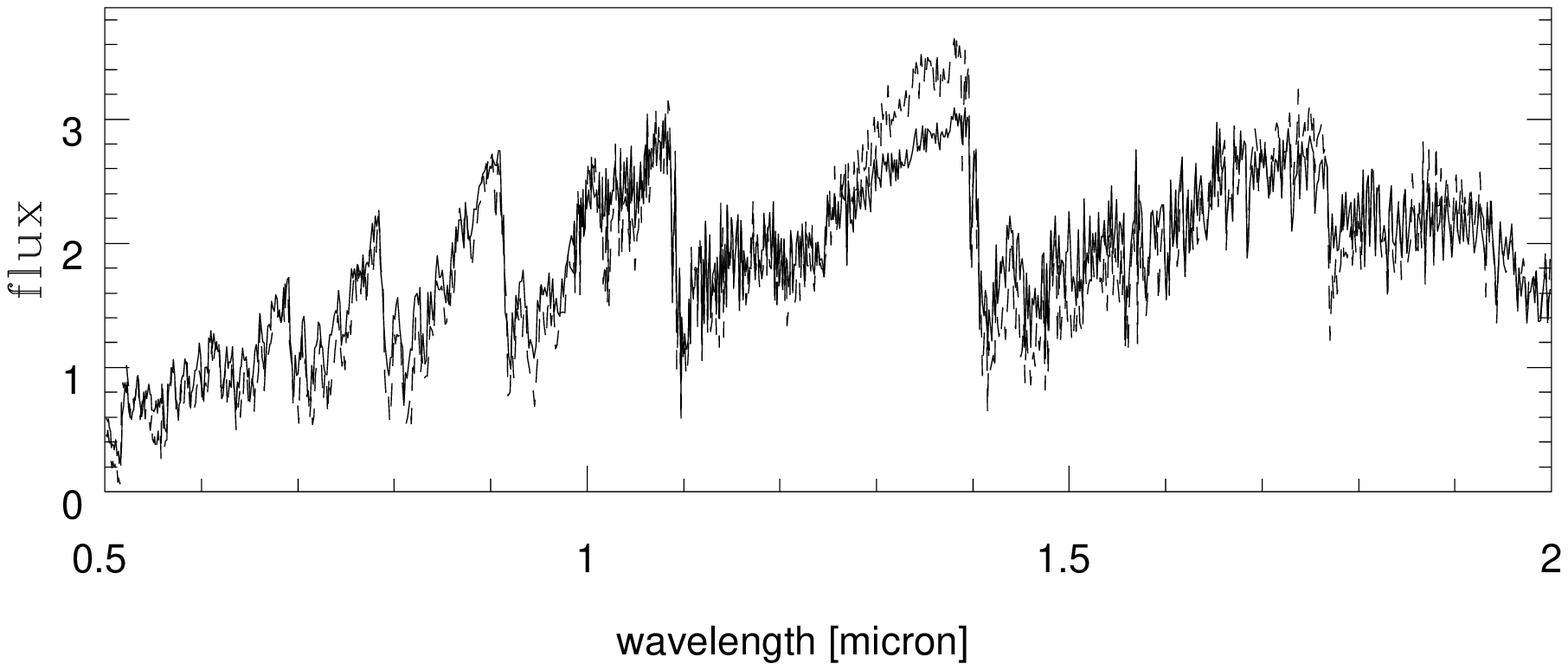}
\caption{\label{model.ps} The model spectrum of BH Cru for two
temperatures: 3200 K (drawn) and 2900 K (dashed)
 }
\end{figure*}

The extended atmosphere is described using the hydrostatic atmosphere models
of \citet{LLJ01}. These models include equilibrium chemistry for a large
number of molecules (but excluding e.g., Mg and Fe molecules and CaCl).
Opacities are included for 7 carbon-star molecules only: CO, CN, CH, C$_2$,
HCN, C$_2$H$_2$ and C$_3$. C/O$=1.01$ is used. The models are shown in Fig.
\ref{atmo.ps} for two different temperatures, corresponding to two different
phases of the pulsation cycle.  Comparison between the two temperatures
confirms the strong dependence of the oxides and C$_2$ on temperature. At the
higher temperatures, C$_2$ is mainly present near the photosphere. At the
lower temperature, the abundance in the extended atmosphere is much higher.
Model spectra at the two temperatures are shown in Fig.~\ref{model.ps}.

\subsection{Optical spectra}

\begin{figure*}
\includegraphics[width=17cm,clip=true]{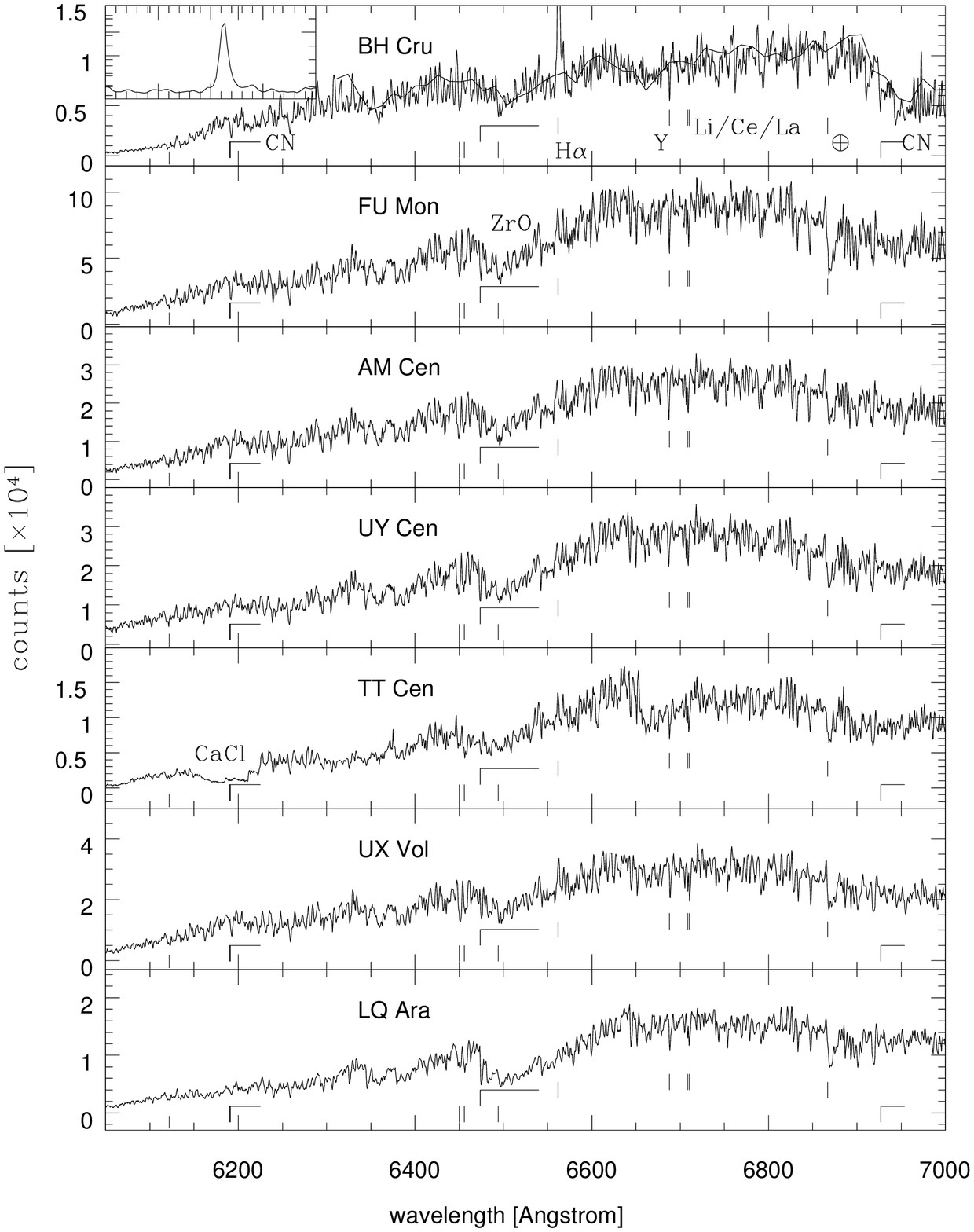}
\caption{\label{spectra.ps} Optical spectra of seven
SC/CS stars. The inset shows the (saturated) H$\alpha$ emission line
of BH~Cru. The spectra are not response-corrected or
flux-calibrated. The strong band at 6857\AA\ is telluric. The spectrum of
BH Cru is shown together with a model for 3200\,K. The y-axis gives total 
counts $\times 10^{-4}$ per wavelength pixel; the noise is negligible,
and all line features are real. A few lines/bands are labeled; indicated
but unlabeled lines/blends are due to CN. }
\end{figure*}

The spectra of oxygen-rich Miras variables show few atomic lines due to the
absorbing molecular blanket which hides the photosphere.  In SC stars, this
blanket is transparent and the spectrum is covered in lines \citep{Tsuji1964}.

Infared spectra of BH Cru are presented by \citet{LLJ01}: they show strong CN
and weak C$_2$ bands.  We obtained optical spectra of BH Cru and some
comparison SC/CS stars using the double-beam spectrograph at the MSSSO 2.3-m
telescope, centred at the 6500\AA\ ZrO band.  The data were taken on 2003 May
18 and 21, under non-photometric conditions.  The dispersion is 0.55\AA/pixel
and the resolution is about 1\AA.  The spectra are flat-fielded using an
internal lamp and sky-subtracted. We did not correct for the atmospheric and
instrumental response.  

The spectra are shown in Fig.~\ref{spectra.ps}.  Typical noise per wavelength
pixel is 1\%. The many lines repeat between the spectra, and are largely due
to s-enhanced metals \citep[e.g.][]{AW98}. The 6708\AA\ line is present in all
other spectra but weak or absent in BH~Cru.  In all cases it is weaker than
the adjacent La{\sc I} 6710\AA. The lithium line overlaps with another line
and the identification is not unambiguous \citep{RWBQ2002}.

The model spectrum of BH Cru (see previous section) is shown in the top
diagram. The model is for 3200\,K: the 2900\,K model also gives an acceptable
fit. The good fit obtained around 6500\AA\ confirms the absence of the ZrO
band (6474--6540\AA) in BH Cru: the good fit is obtained in spite of this band
not being included in the opacities in the model.  All other stars which were
observed show the band, although with varying strength.  The broad 6350\AA\ 
ZrO band is similarly seen in all stars except BH Cru.

BH Cru shows possible absorption lines at 6053 \&\ 6070\AA, which
differ from the other stars and may be from the C$_2$ Swan-band. Other lines
of the Swan bands at 6190 \&\ 6480\AA\ are not detected, but the
confusion at our resolution is too high to rule out their presence.
C$_2$ is present in BH Cru longward of 7000\AA\ \citep{LLJ01}.

\begin{figure}
\includegraphics[width=8.45cm,clip=true]{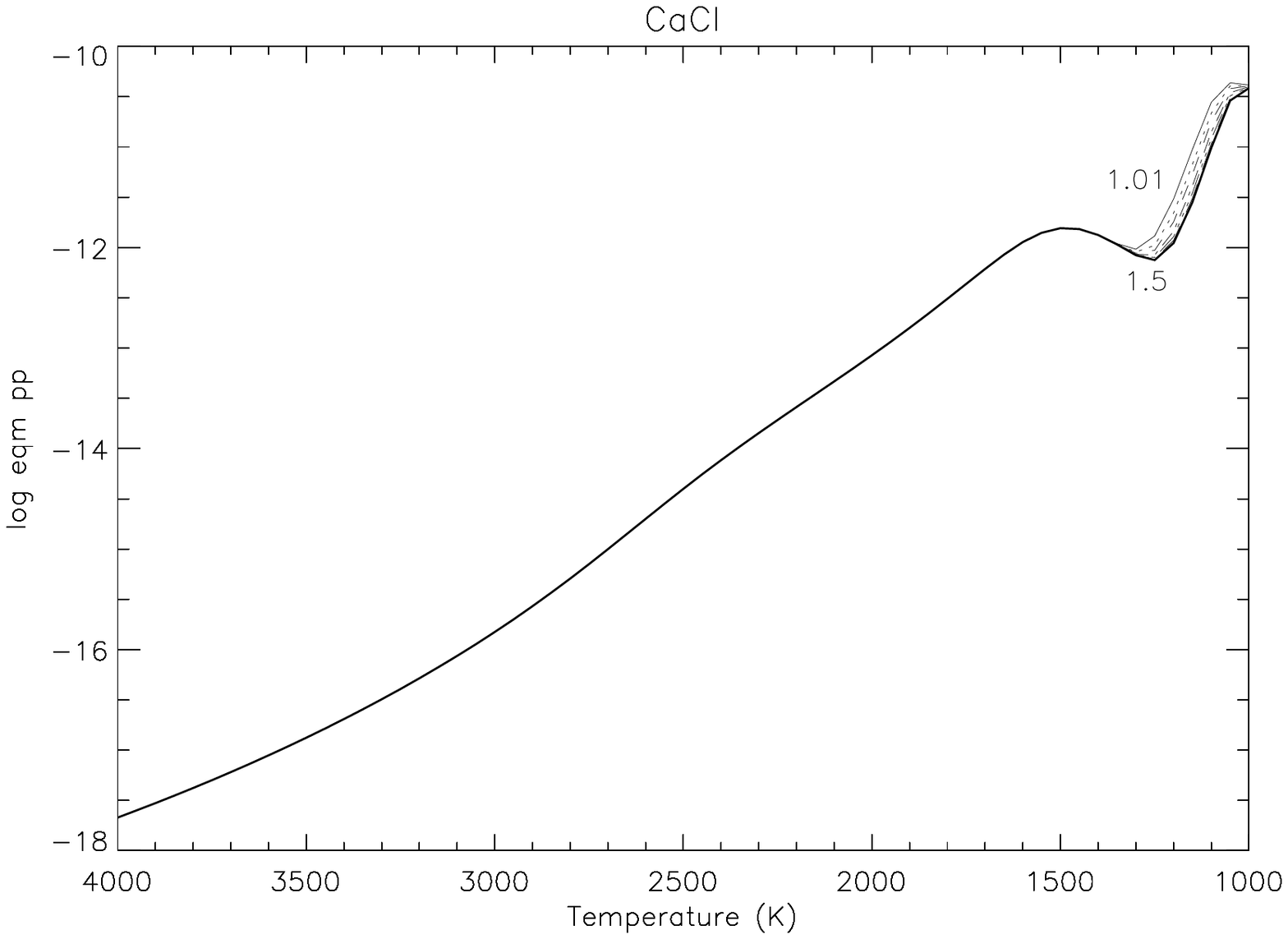}
\caption{\label{cacl.ps} Same as Fig.~\ref{c2.ps}, but for CaCl}
\end{figure}

The double CaCl band at 6200\AA\ is seen in TT Cen only. This band is common
in carbon stars, but is often blended and masked by strong CN bands.
The CaCl abundance depends very little on the C/O ratio (Fig. \ref{cacl.ps}).
The CaCl band strength tends to vary over the pulsation cycle
\citep{Sanford1950}, being strongest near minimum light: TT Cen was observed
near minimum. TT Cen may be compared to VX Aql \citep{WW76}, in which CaCl
also is the strongest molecular band.  Both these stars are long-period Miras.
\citet{CW77} show that CaCl is favoured by high density environments.

\subsection{Dust}

Stars with C/O close to unity have difficulty forming dust: the removal of C
and O from the chemistry means that neither silicates nor carbon grains can
form. It is therefore not surprising that the SC stars show little evidence
for significant dust emission. A few ISO spectra have
very poor S/N at mid-infrared wavelengths, and weak features would not
be seen.

We therefore retrieved the IRAS LRS spectra of the seven SC/CS stars
with 12-$\mu$m fluxes above 15\,Jy: BH~Cru, AM Cen, FU Mon,
UY Cen, S Lyr, VX Aql, and WZ Cas.  The spectra mostly show adequate
but low S/N (the brightest star in the sample, UY Cen, only has
$F_{12}=55\,$Jy).  In order to emphasize any excess, we have
subtracted a pseudo-continuum represented by a $\lambda^{-2}$ power
law that was scaled to the flux at 14\,$\mu$m for each star.

\begin{figure}
\includegraphics[width=8.45cm,clip=true]{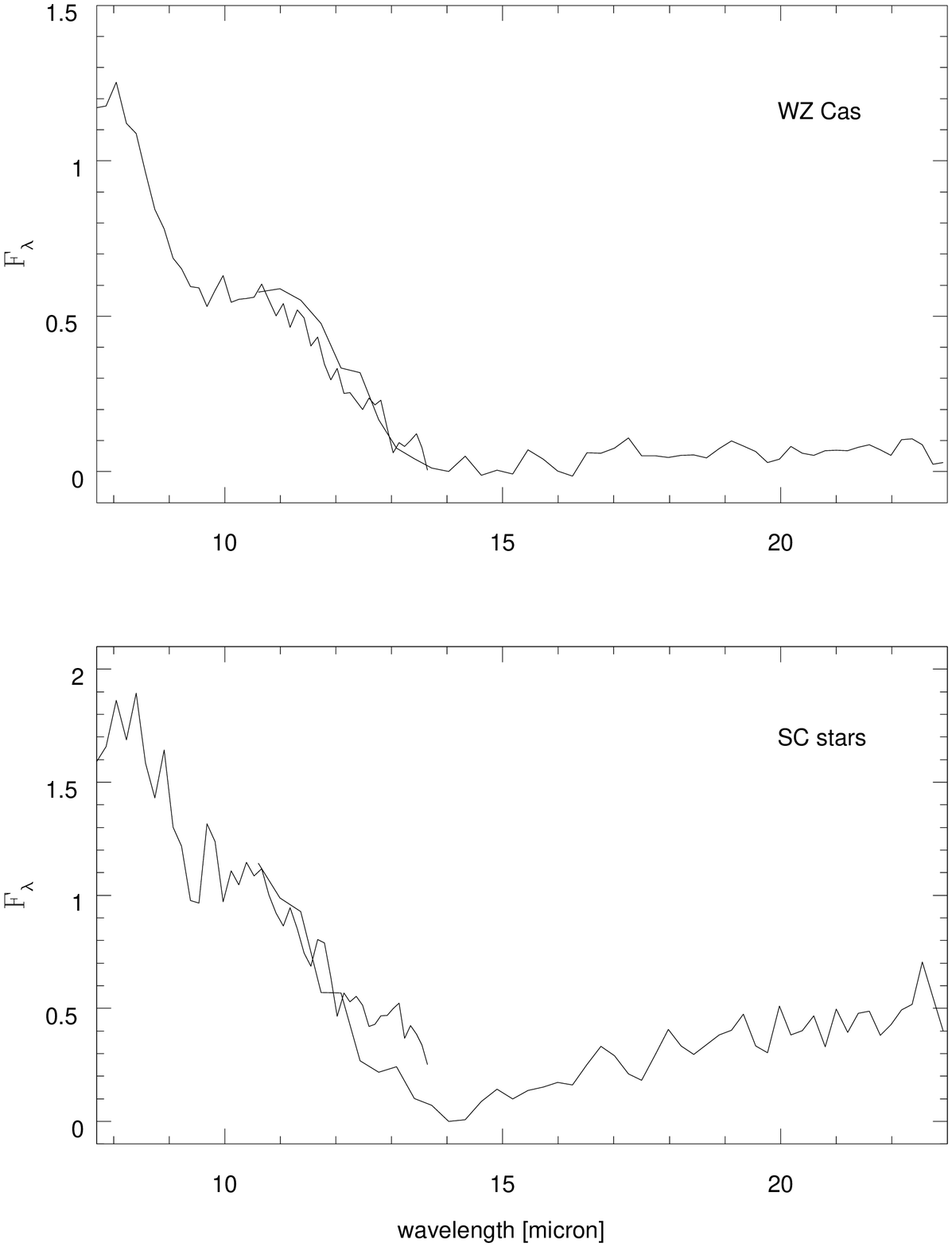}
\caption{\label{lrs} IRAS LRS spectra of SC/CS stars, from which a
$\lambda^{-2}$ pseudo-continuum has been subtracted. Top panel: WZ Cas, the
only star showing a silicate-like feature. Bottom panel: average of 
six other SC
stars.  }
\end{figure}

The top panel of Fig.~\ref{lrs} shows the resulting spectrum of WZ
Cas.  The impression of a weak silicate feature is deceptive:  it reflects 
a deep CS absorption feature at 8$\mu$m.

 None of the other SC LRS stars resembles WZ Cas.  We have co-added
the remaining six spectra and again subtracted a pseudo-continuum to
produce a 'typical' SC-star LRS spectrum (lower panel of
Fig.~\ref{lrs}). Compared to WZ Cas, the continuum-subtracted spectrum
rises beyond 15 $\mu$m.  The wavelength coverage is insufficient to
decide whether this is a dust continuum or a broad feature.  However,
the rise resembles that seen in some young stellar objects, where it
is part of a feature peaking at 23$\mu$m. This feature has previously
been attributed to FeO, but \citet{KHB2002} found a fit to the 
iron-sulfide troilite.  \citet{HBK2002} report the same feature and
presumably the same mineral in planetary nebulae, and suggest a
carbon-star origin.  If the SC stars show the same feature, it
would strongly support the identification with troilite, as 
formation of this mineral should benefit from the removal of O from the
dust formation.

\begin{figure}
\includegraphics[width=8.45cm,clip=true]{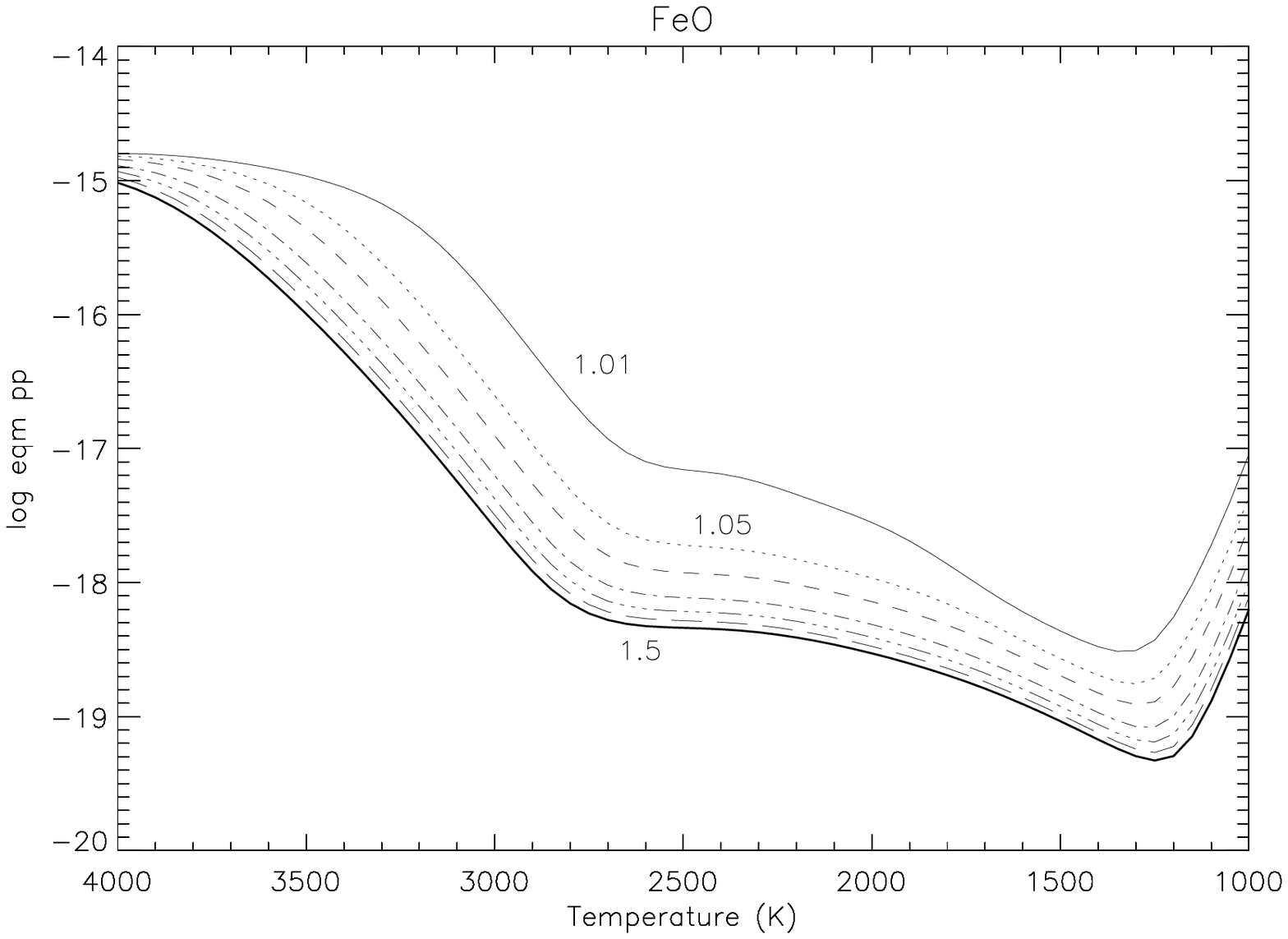}
\caption{\label{feo.ps}  Same as Fig.~\ref{c2.ps}, but for FeO}
\end{figure}

\begin{figure}
\includegraphics[width=8.45cm,clip=true]{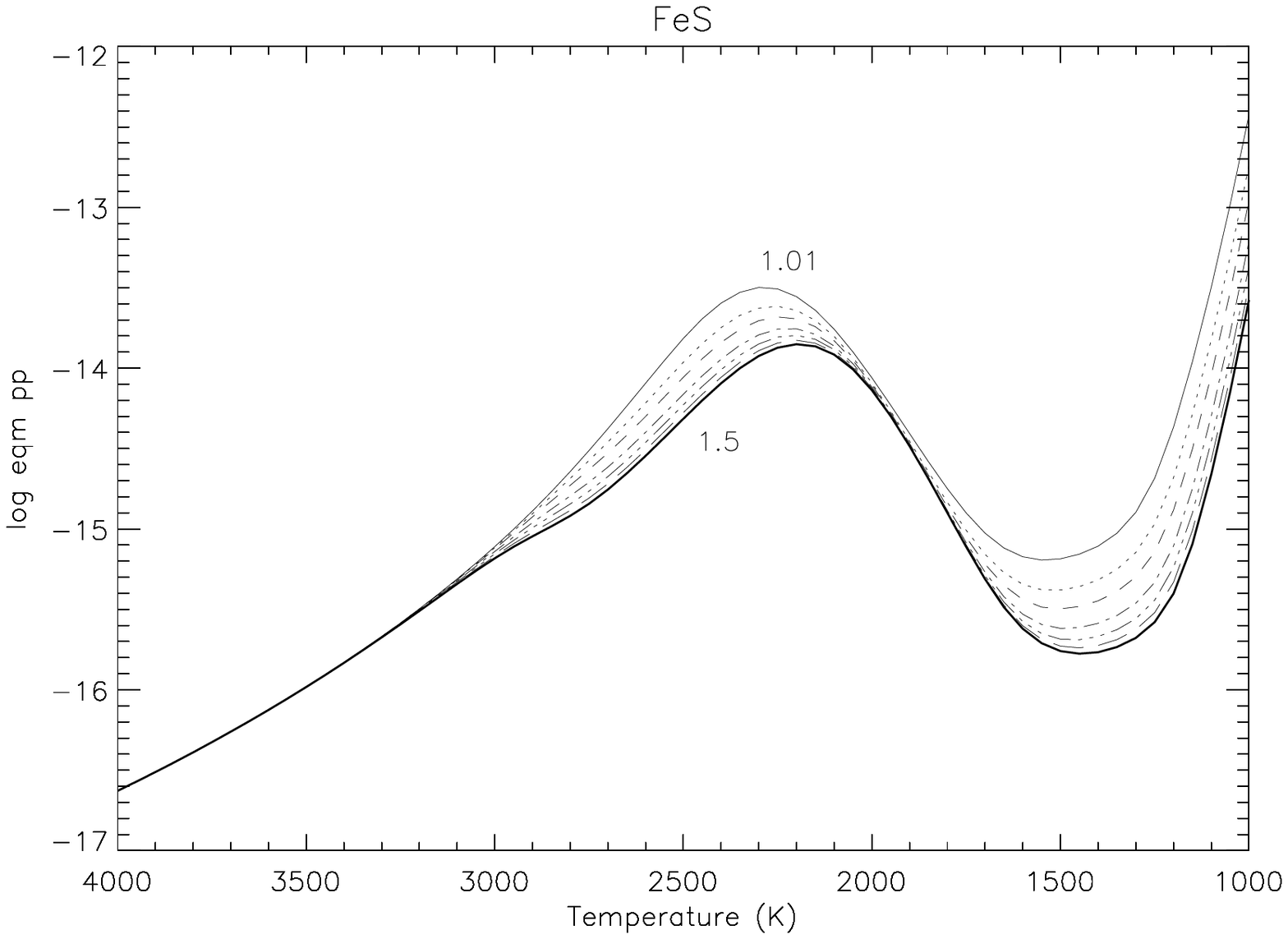}
\caption{\label{fes.ps}  Same as Fig.~\ref{c2.ps}, but for FeS}
\end{figure}

Figures \ref{feo.ps} and \ref{fes.ps} show the calculated equilibrium
abundances of FeO and FeS\@. As expected, FeO
declines sharply as the C/O ratio increases. FeS also shows a decline but
is less sensitive especially at photospheric temperatures. At all but the
lowest temperatures, FeS is the more abundant molecule by three orders of
magnitude. The formation of an iron sulfide such as troilite would
therefore not be unexpected. However, the evidence for a troilite feature
is weak and this point remains to be confirmed.

\citet{FGDO2000} suggest that FeSi would be the main dust component in S
stars. This mineral has bands at 32 and 50 micron. ISO SWS spectra are
available for three SC stars: WZ Cas, S Lyr and W Cas. The S/N is low and no
emission can be seen longward of 30 micron.

Evidence for mass loss in BH Cru comes from a blue-shifted (by $19\,\rm
km\,s^{-1}$) possible rubidium line \citet{AW98}.  Otherwise, the SC stars
show little evidence for significant mass loss. (Of the stars in Table
\ref{sc_cs.dat}, only FU Mon was detected in CO and a mass-loss rate of
$10^{-7}\,\rm M_\odot\,yr^{-1}$ derived with a very low expansion velocity
($2.8 \,\rm km\, s^{-1}$) \citep{JK98}.)  This is in marked contrast to the
oxygen-rich Miras and carbon Miras, even though the SC stars form an
intermediate evolution.  It is however understandable, as the mass loss is
driven by radiation pressure on dust. SC stars are expected to form a phase of
mass-loss interruption \citep{ZLWJ92} while evolving from an oxygen-rich to a
carbon-rich star.

\section{LX~Cyg: another BH Cru}  

Large period changes similar to BH Cru are very rare among Mira variables:
\citet{ZB03} describe 6 cases, including R Hya. They estimate that the type of
evolution seen in BH Cru occurs in $\sim 1$ per cent of well observed Miras,
over a century of observations.  To find one on-going case among the small
group of SC/CS stars (16 in total) would be unexpected.  To find a second one
would suggest a correlation between the rare spectral class and the period
evolution.

There is in fact a second case of period evolution among the SC stars.  LX Cyg
is an SC3 star with a catalogued (GCVS) period of 465 days.  An increase of
its period was independently discovered by \citet{AA85} and by
\citet{Broens2000} and \citet{TMP2003}.  \citet{TMP2003} have analyzed an
extended data set.  They find that historical records of LX~Cyg show the
period was fairly stable at about 460d from its first measurement around JD
2426000 to 2439000, after which it increased to around 580 days.  We note that
BH Cru and LX Cyg are the two SC stars in Table \ref{sc_cs.dat} with the
latest subtypes and longest periods. A spectrum kindly taken for us by James
Bryan in 2003 confirms that LX Cyg is an SC star, but does not allow a current
subclass to be determined.

\begin{figure*}
\includegraphics[width=0.8\textwidth, clip=true]{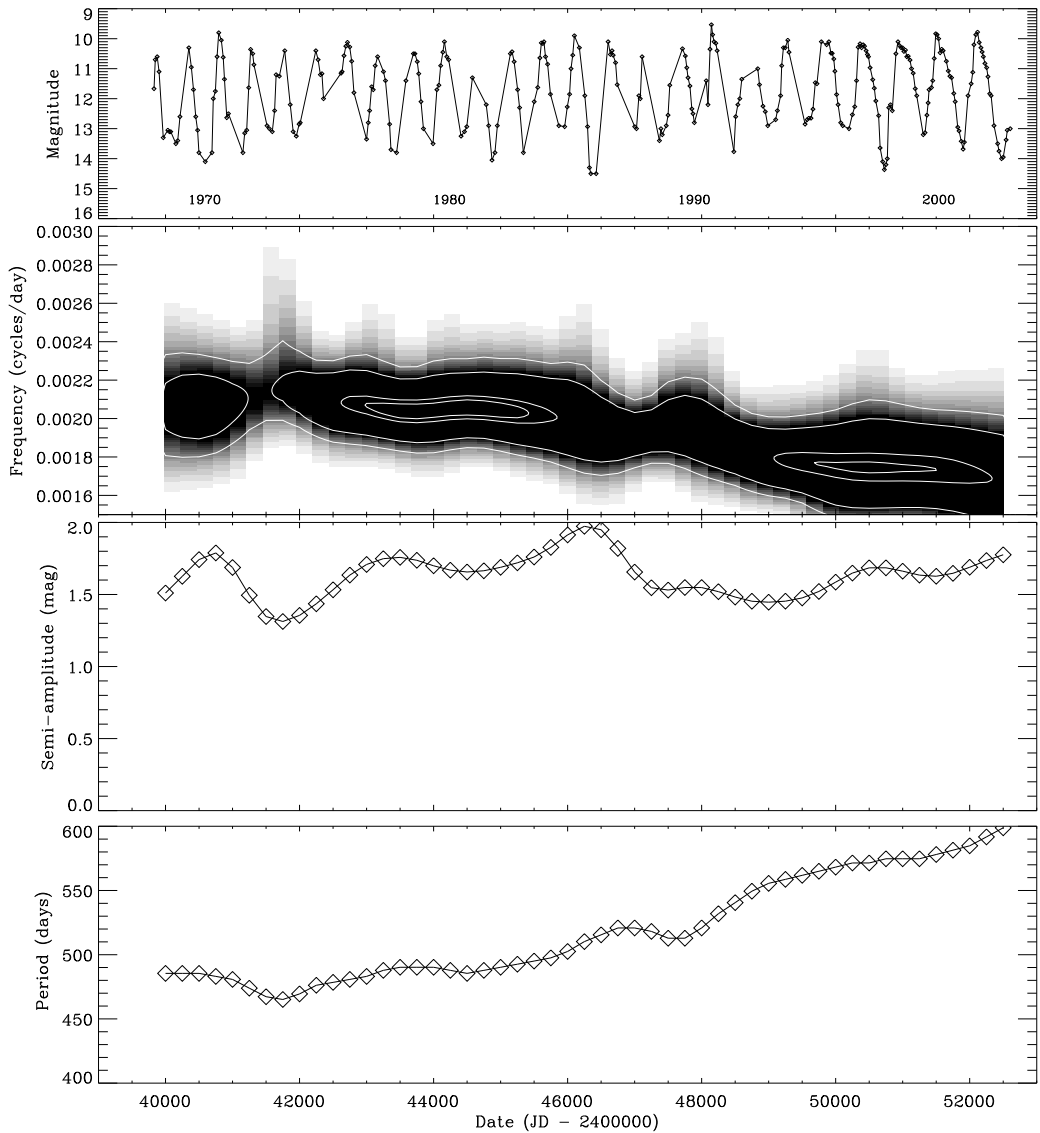}
\caption{\label{lxcyg.ps} The wavelet analysis for LX~Cyg. The lower three
panels are on the same scale as Fig.~\ref{bhcru.ps}, to facilitate
comparison with BH~Cru}
\end{figure*}

We carried out a wavelet analysis of the AAVSO data, identical to the one
shown for BH Cru. The result is shown in Fig.~\ref{lxcyg.ps}: the axis scales
are the same as in Fig.~\ref{bhcru.ps}, to facilitate the comparison.  The
rate of period change in LX~Cyg is very similar to that in BH~Cru. The
pulsation amplitude is somewhat larger and has not changed significantly, and
the total change in period is larger.

A redetermination, or in several cases a first determination, of the periods
of the SC/CS stars is given Appendix \ref{appA}. There we show that our period
of VX Aql (609\,days; unknown subtype) is inconsistent with a earlier
determination (470\,days).  We have insufficient data to confirm a period
cange, but it seems possible that VX Aql is a third case of a SC star with an
evolving period. If true, {\it all} SC star with periods longer than about 470
days show this effect.  Many normal Miras with such periods also show
fluctuations \citep{ZB03}, but limited to about 10 per cent in period. The SC
stars show much larger changes. Even without VX Aql, a link between the
C/O ratio near unity, and the unstable periods, seems likely.

\section{Discussion}
\label{origin}

A few Miras are known to show changing periods.  The well-known case of R Hya
was studied by \citet{ZBM02} who find evidence for a stable period of 495 days
before AD 1800, followed by a declining period, and a newly stable period of
385 dats since AD 1950. R Aql shows similar behaviour. TY Cas was recently
added to the list of variable periods \citep{HM03}.  But overall, period
evolution among Miras is rare.  \citet{GH00} examined a sample of 100 Miras,
and apart from the previously known case of T UMi \citep{GS95a,MF95,SKB03},
found no further case of long-term period change.  \citet{Woo76} studied 45
stars and found small period changes with greater than 99 per cent
significance in 6 cases, but no large-scale evolution.  \citet{MF2000}
performed a trend analysis of 383 long period variables using 90 years of
AAVSO data. Only 9\%\ showed evidence for trends, and of these only 9 stars
showed strong trends.  \citet{PA99} analysed a sample of 391 Mira variables
and found some evidence for a slow average increase in periods with time, but
few or no stars with large period variability.

This suggests that over the past 50--100 years, period evolution has only
occurred in $\sim 1\%$ of well-observed Miras. To find a case among the seven
SC/CS stars classified in the GCVS as Miras with established periods is not
ruled out by this rate. But the discovery of a second case implies an
overrepresentation of the SC/CS stars.  The rate of occurence of period
evolution among semi-regular variables is not known.

Period changes in Miras have normally been attributed to a recent thermal
pulse or helium flash \citep{WZ81}. This explains the rarity of such objects.
Fast period evolution will only occur during and immediately after the helium
flash \citep{VW93}.   At any one time, only $\sim 0.1\%$ may be found at the
rapid period increase. This model would therefore suggest that BH~Cru is in a
unique evolutionary phase, following a thermal pulse occuring during the
1960s.

But the fact that, among the small number of SC stars, LX Cyg is also
undergoing similar changes, makes the thermal-pulse explanation rather
unlikely.  The fact that historical records suggest recurrent period changes
in BH Cru also argues against a thermal pulse.  There are alternative models
for period changes, predicting a non-linear instability in the Mira structure.
In the case of R Hya \citep{ZBM02} those may yield a better fit to the data.

In AGB stars, the visual pulsation amplitude is largely determined by varying
molecular opacities: e.g., the extreme amplitudes of oxygen-rich Miras is
caused by the varying TiO and VO bands.  The low molecular opacities in SC
stars can explain their low pulsation amplitudes (Appendix
\ref{amplitude}). However, the molecular abundances become very sensitive to
small changes in the chemical equilibrium, and this may affect the pulsation
amplitude.  An increased pulsation will lower the effective temperature and
this shifts the chemical equilibrium.  This process can result in a positive
feedback, which leads to larger pulsation amplitudes.  \citet*{BCZ2000} have
suggested that a change in the amplitude can cause a period change (rather than
the usually assumed reverse relation) due to non-linear effects.
The positive feedback on the amplitude may thus affect
the period as well, and may be a possible cause of the pulsation
instability observed in BH Cru, LX Cyg and possibly VX Aql.

Finally, although we argue against a thermal pulse, we note even if a thermal
pulse did occur, the C/O ratio in BH Cru would not have changed.  Carbon
dredge-up following a thermal pulse does not commence until $\sim 250\,$yr
after the thermal pulse \citep{Mow99}, when the radius has returned to its
pre-thermal pulse value, and is continuing to decrease.  The dredge-up for the
models of \citet{Mow99} lasts for about 100\,yr while the period of the star
is {\it de}creasing, well after the initial sharp period increase. Thus, in
either case, the change from SC to CS star would be due to a lower
temperature, and not a increase in the C/O ratio. The distinction between SC
and CS stars may measure effective temperature, rather than an evolutionary
progression \citep{S73}.

\section{Conclusions}

Our investigation of the evolving variable BH~Cru shows an increase in
its period of 25\%\ within 25\,yr. The period has stabilized at
about 540 days. The visual semi-amplitude has increased simultaneous
with the period increase, up to a value of 1.25\,mag.  The changing
period shows that the radius of the star has increased, and the
temperature decreased, the latter confirmed by a slow reddening.

The spectral type of BH Cru changed simultaneously from SC to CS.  Chemical
equilibrium modelling, both for the potosphere and a hydrostatic atmosphere,
explains this as due to the decrease in the effective temperature. The lower
temperature favours formation of C$_2$ and causes the fractional abundances of
ZrO and YO to decrease.  Our calculations explore a range of C/O$>1$, for
which oxides are still found although at reduced abundances.  The distinction
between SC and CS stars does not require an evolution from C/O$<1
\longrightarrow >1$, as sometimes is suggested, although this can also play a
role.  Infrared spectra suggest the possible presence of the iron-sulfide
troilite in SC stars. Chemical calculations show that FeS is abundant in the
photosphere, and in the absence of silicates or carbon grains, may be an
important dust component.

We determined new periods for a number of SC/CS stars, including three
for stars with no previously known period. One star, LX~Cyg, shows a
much longer period than previously determined, and is confirmed as
undergoing period evolution similar to BH~Cru.  VX Aql also shows an
inconsistency between the present period and one previously reported.
Among the few Miras known with evolving periods, BH~Cru was unique in
showing an {\it increasing} period. Its time scale for the evolution is ten
times faster than that of the well-studied case of R Hya.  With LX~Cyg, a
second case of rapid period increase is now known.  

Period changes in Miras are commonly attributed to thermal pulses.  This
appears unlikely in BH Cru, because other SC stars show the same type of
changes.  As an alternative model we suggest the possiblity that a feedback
between molecular opacities, pulsation amplitude and periods cause unstable
periods among the long-period SC stars.

\section*{Acknowledgements}

We appreciate the efforts of Ranald McIntosh in maintaining the RASNZ
database, and of Brendon Brewer, who worked on some of the visual data as a
vacation student.  TRB is grateful to the Australian Research Council for
financial support, and AAZ, MM and AJM acknowledge a PPARC rolling
grant. PPARC also supported this research via a visitor grant.

\bibliographystyle{mn2e}

\bibliography{paper-bhcru}

\appendix

\section{Periods of the other SC stars}
\label{appA}

\begin{table}
\caption[]{\label{sc_cs.dat} GCVS data for the SC and CS stars.
Stars in the last section are classified as SC by \citet{CF71} and
\citet{KB80}: the GCVS gives 
these as C stars.
  }
\begin{flushleft}
\begin{tabular}{llllll}  
\hline
  name & var. & $m_V$ & Period & Period  & spectral type \\
       & type & (max) & (GCVS)   & (this work)  & \\
       &      &       & [days] & [days]    \\
\hline
   LQ    Ara & M:   & 10.3 &      & 193? & SC              \\
   AM    Cen & LB   & 10.4 &      & 276? & SC              \\
   V372  Mon & SR   & 12.5 &      &     & SC(N)           \\
   V3832 Sgr & LB:  & 13.4 &      &     & SC              \\
   UX    Vol$^a$ & LB:  & 9.2  &      & 182 & SC              \\
   BH    Cru &  M   & 7.2  &  540 & 430--530    & SC4.5/8-SC7/8 \\
   UY    Cen & SR   & 9.22 &  115 &     & SC              \\
   CM    Cyg & M    & 9.3  &  255 &     & SC2-S4e         \\
   LX    Cyg & M    & 11.5 &  465 & 470--600 & SC3e-S5,5e:     \\
   S     Lyr & M    & 9.8  &  438 & 436 & SCe             \\
   GP    Ori & SRB  & 12.2 &  370 & 184, 339 & C8,0J:(SC)ea    \\
\hline
   CY    Cyg &  LB  & 11.0 &      & irregular & CS(M2p)         \\
   R     CMi &  M   & 7.25 &  338 & 337 & C7,1Je(CSep)    \\
   TT    Cen &  M   & 11.5 &  462 &     & CSe  \\
   FU    Mon &  SR  & 11.6 &  310 &     & C8,0J(CSe) \\
   RZ    Peg &  M   & 7.6  &  439 & 437 & C9,1e(Ne)/CSe \\
\hline
{\it   VY    Aps} & SRA  & 9.6  &  152 &       &    \\
{\it   AM    Car} & SR   & 14.3 &  314 &       &    \\
{\it   R     Ori} & M    & 9.05 &  377 &       &    \\
{\it   RR    Her} & SRB  & 8.8  &  240 &       &    \\
{\it   W     Cas} & M    & 7.8  &  406 &       &    \\
{\it   WZ    Cas} & SRa  & 9.4  &  186 &       &    \\
{\it   VX    Aql} & M:   &      &      & 604   &    \\
\hline \\
\end{tabular}
$^a$
We note that the GCVS gives the wrong declination for this star.  The
correct values are $\alpha,\delta = 08^{\rm h} 46^{\rm m} 17.6^{\rm s},$ 
$-71^{\rm o} 02^\prime 20^{\prime\prime}$ (J2000)
\end{flushleft}
\end{table}

The General Catalogue of Variable Stars (GCVS) lists only 11 SC stars
(including BH~Cru) and 5 CS stars.  These are listed in
Table~\ref{sc_cs.dat}.  \citet{KB80} and \citet{CF71} list a few other SC
stars that the GCVS gives as carbon stars, and which perhaps should be
classified as CS.  In Table~\ref{sc_cs.dat}, these stars are in the third
group (in italics).

\label{amplitude}

Of the 16 known SC and CS stars, 6 have no catalogued period.  We attempted
to determine some of their periods, and to confirm existing periods,
using new data (see below) and data from amateur databases.  New
light curves and/or power spectra were determined for the six stars,
as follows.

\paragraph*{LQ~Ara, AM~Cen, UX~Vol:}

\begin{figure}
\includegraphics[width=8.45cm,clip=true]{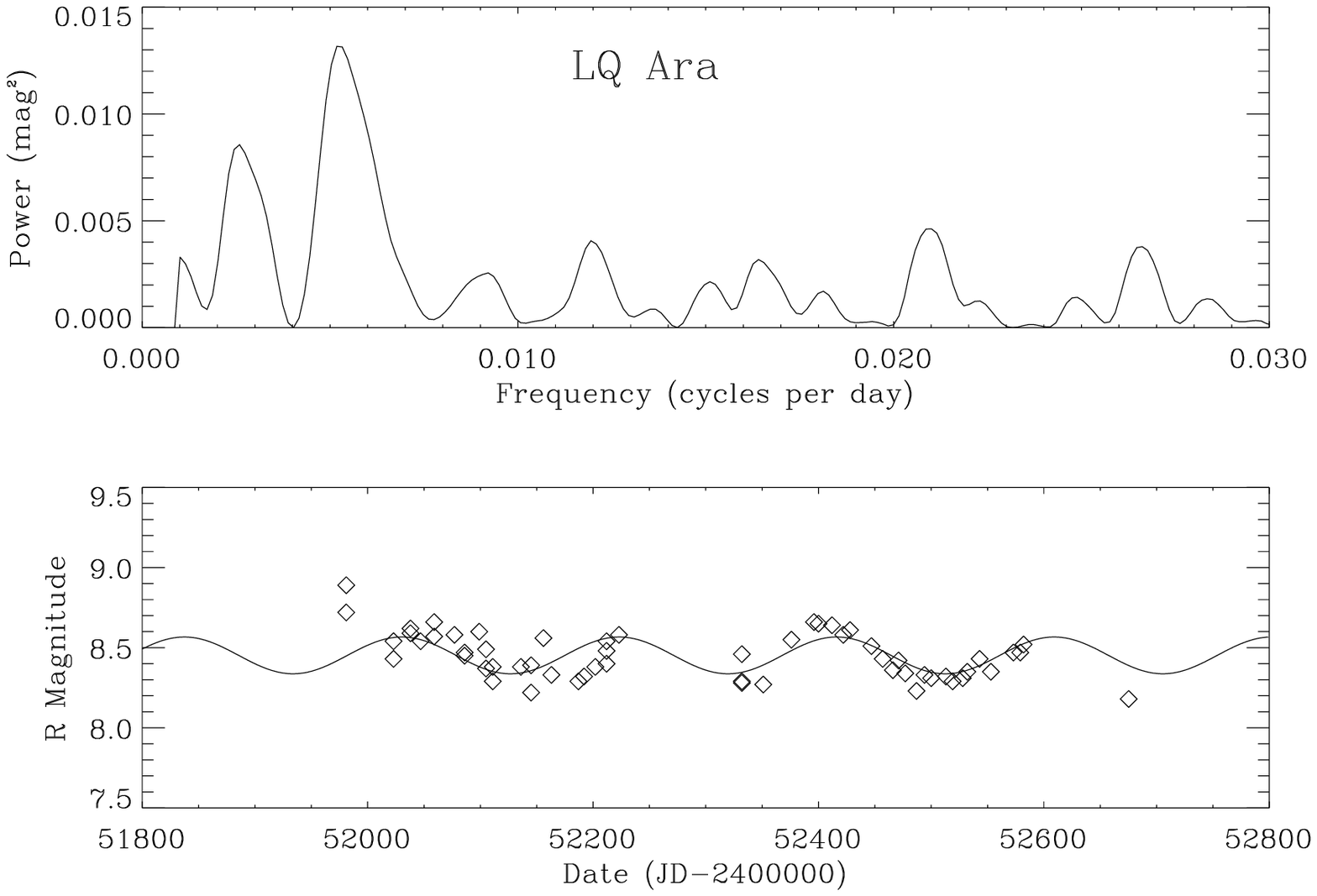}
\caption{\label{lqara.ps} Power spectrum (upper panel) of the CCD light
curve (lower panel) of LQ Ara; the solid curve shows the best-fitting 
sinusoid.  }
\end{figure}

\begin{figure}
\includegraphics[width=8.45cm,clip=true]{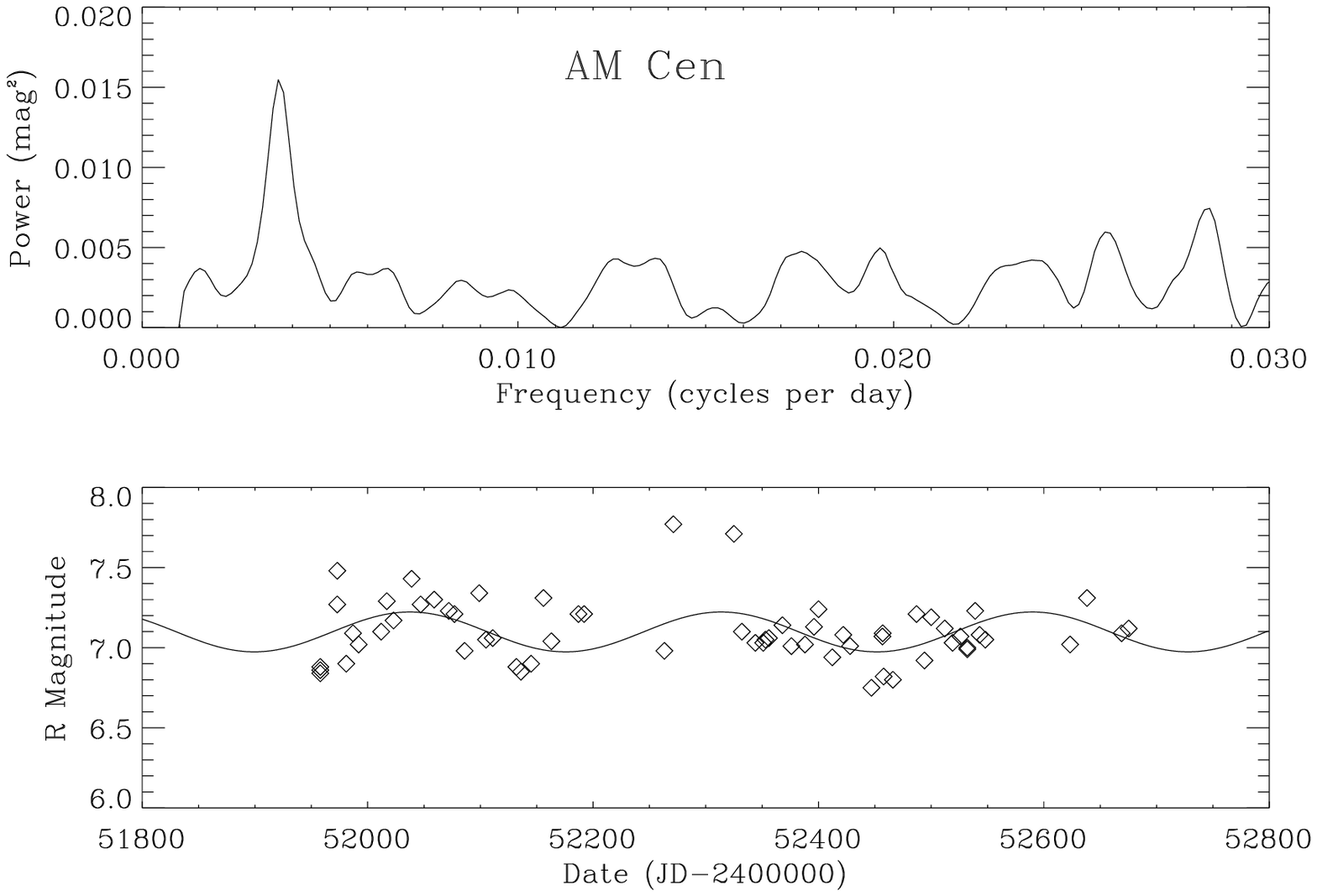}
\caption{\label{amcen.ps} Power spectrum (upper panel) of the CCD light
curve (lower panel) of AM Cen; the solid curve shows the best-fitting 
sinusoid.  }
\end{figure}

\begin{figure}
\includegraphics[width=8.45cm,clip=true]{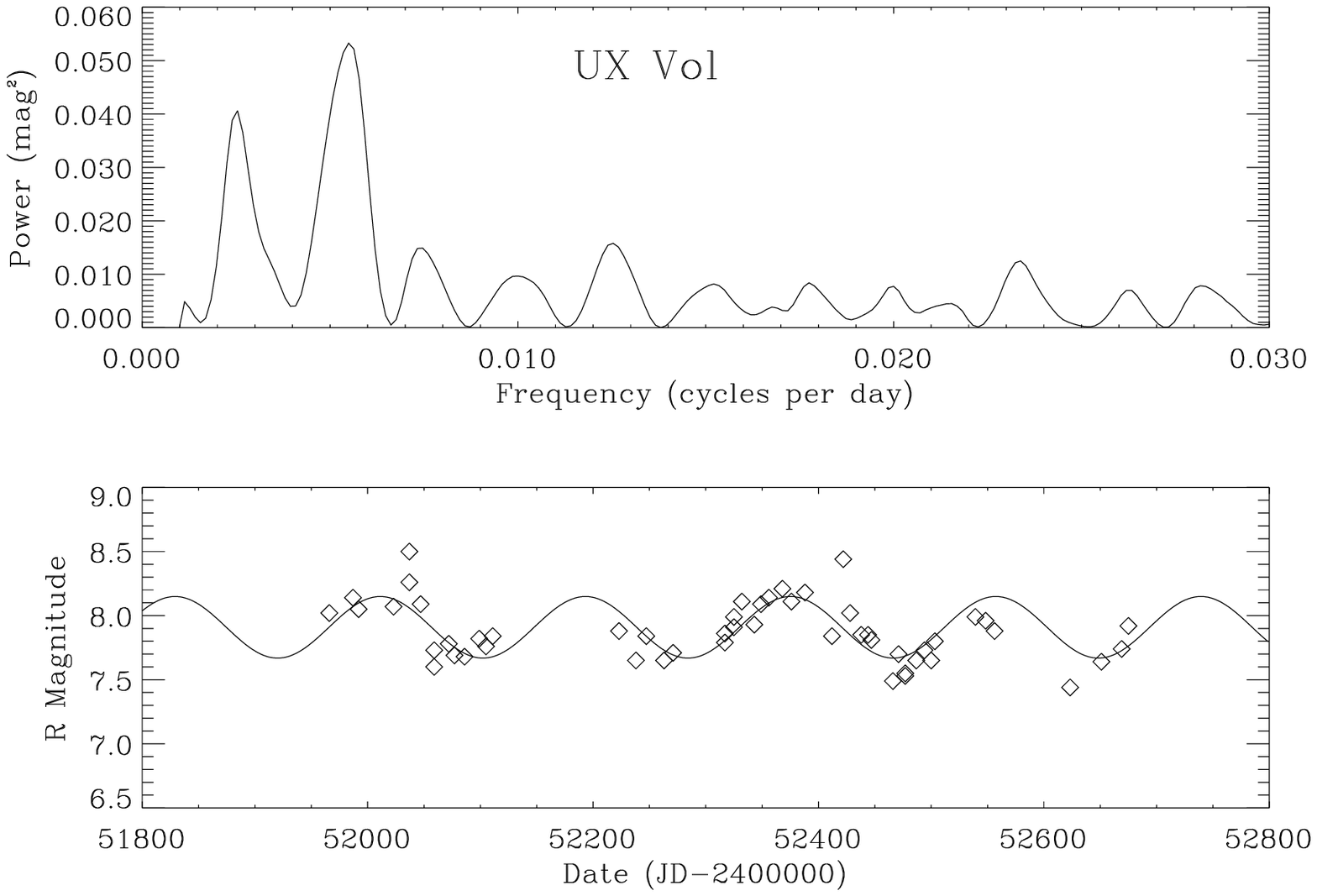}
\caption{\label{uxvol.ps} Power spectrum (upper panel) of the CCD light
curve (lower panel) of UX~Vol; the solid curve shows the best-fitting 
sinusoid.  }
\end{figure}

Photometric observations of these three stars were obtained by one of us
(VT) using CCD imaging with an unfiltered 100mm telephoto lens.  For
LQ~Ara, the light curve and power spectrum (Fig.~\ref{lqara.ps}) indicate a
period of 193 days.  The amplitude is low, indicating that this star is a
semiregular (not a Mira, as suggested in the GCVS).  For AM Cen
(Fig.~\ref{amcen.ps}) we find a possible period of 276 days, but this
requires confirmation with more data.  The star UX Vol
(Fig.~\ref{uxvol.ps}) appears to be a semiregular with a period of 182
days.

\begin{figure}
\includegraphics[width=8.45cm,clip=true]{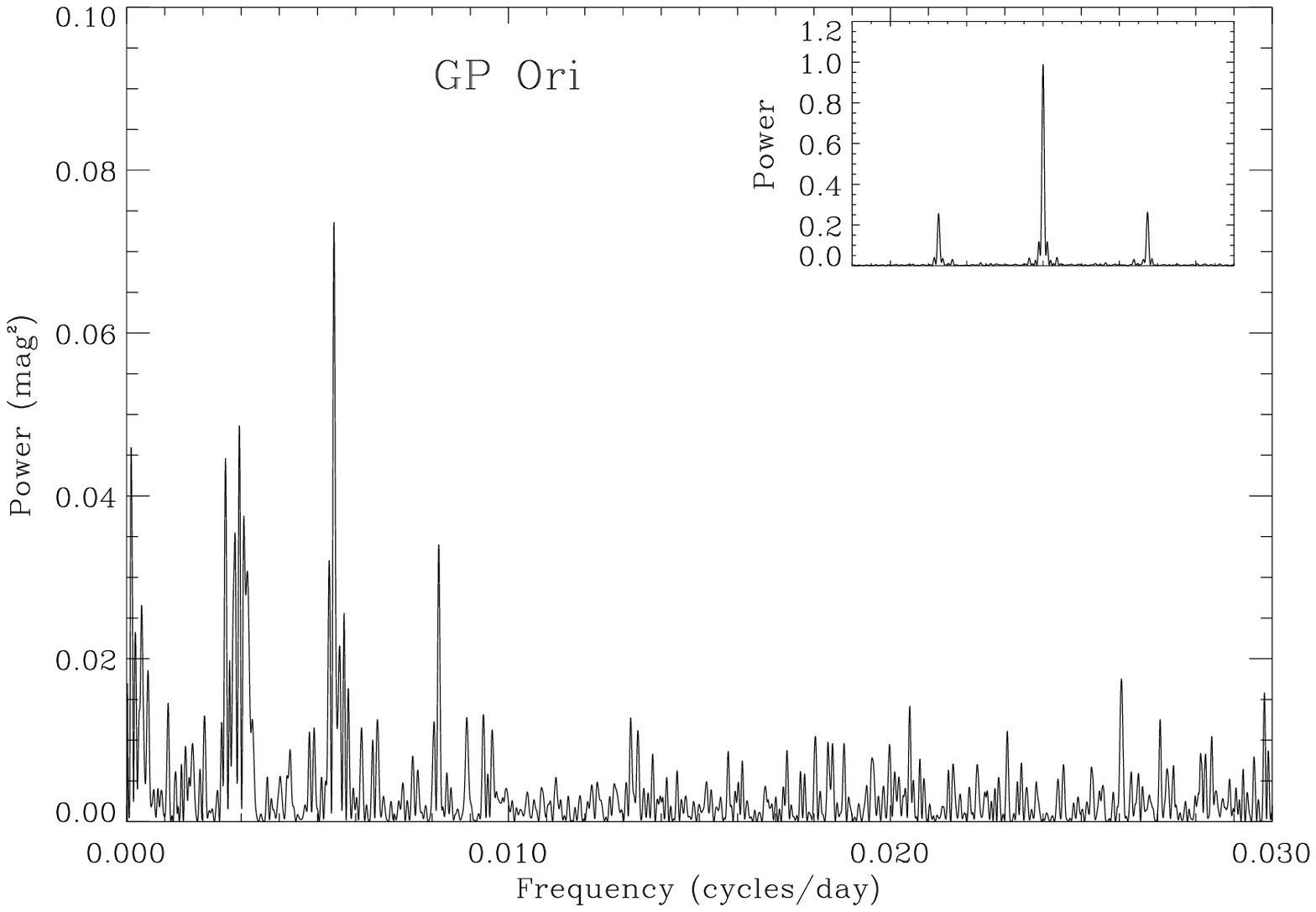}
\caption{\label{gpori.ps} Power spectrum of the light curve
of GP Ori showing two distinct periods.  The inset shows the spectral
window. 
 }
\end{figure}

\paragraph*{GP Ori:} This star is classified as a semiregular (SRb) with
a period of 370\,d in the GCVS.  We have analysed about 440 visual
estimates from the AAVSO and VSOLJ databases, spread rather sparsely
over the past 30 years.  The power spectrum, shown in
Fig.~\ref{gpori.ps}, confirms GP~Ori as a semiregular with two
periods: 184\,d (0.00543\,d$^{-1}$) and 339\,d (0.00295\,d$^{-1}$).
The period ratio of 1.84 is typical of many semiregulars
\citep{KSC99}.  The power spectrum is somewhat complicated by yearly
gaps in the data (the peak at 0.081\,d$^{-1}$ is an alias), and this
is made worse by the fact that the shorter period is very close to
half a year.  Nevertheless, both periods seem secure.  \citet{BZ98}
showed that in double-period semiregulars, the longer period tends to
agree with the Mira period--luminosity relation while the shorter
period falls on the semi-regular sequence (sequences C and B of
\citealt{Woo99}).  We also note that the power spectrum shows broadened
envelopes that have been interpreted by \citet{Bedding2003} as
evidence for solar-like oscillations (i.e., stochastic excitation,
presumably from convection).

\paragraph*{CY Cyg:} We have analyzed 424 AAVSO observations from
JD 2438540 to JD 2452653. The star seems completely irregular, and the Lb
classification (which is usually based on lack of data) seems warranted.

\paragraph*{VX Aql:} 

\begin{figure}
\includegraphics[width=8.45cm,clip=true]{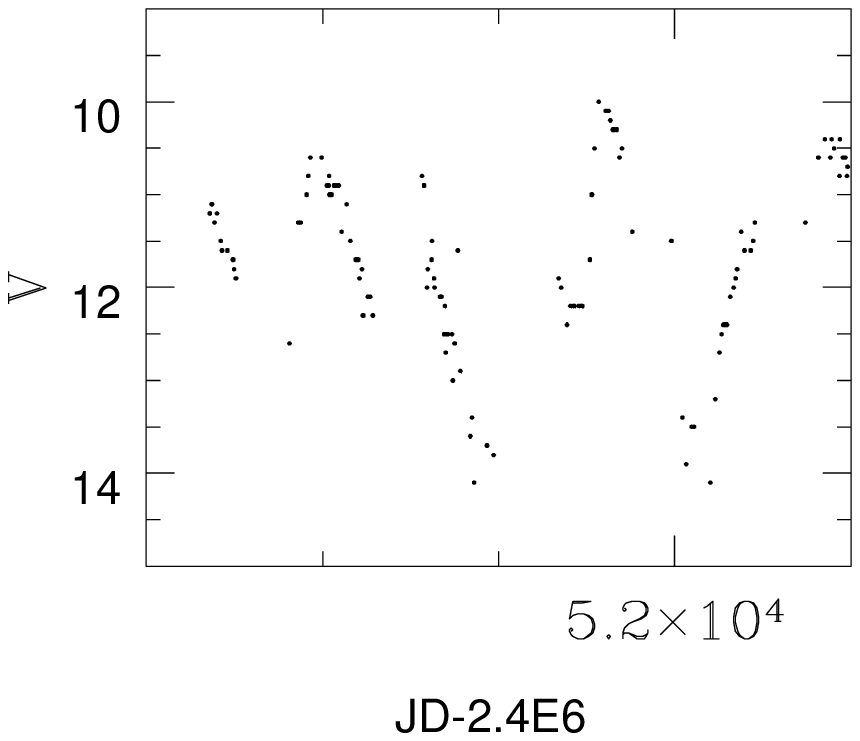}
\caption{\label{vxaql.ps} VSOLJ Light curve of VX Aql 
 }
\end{figure}

This star is of particular interest because of a C/O ratio suggested
to be exactly unity \citep{GW75}.  The GCVS lists it as a Mira but
does not give a period.  A VSOLJ light curve is shown in Fig. \ref{vxaql.ps}.
The AAVSO light curve can be found on the AAVSO web site.
We have analyzed 86 observations from a single AAVSO
observer which show the star is a Mira, with a period of 604 days and
a very pronounced double maximum.  Interestingly, \citet{WW76} adopted
a period of $\sim$470\,d which they attributed to \citet{Kur58}.  It
seems possible that VX~Aql is another example of an SC/CS star that
has shown a large period increase.

\paragraph*{RZ Peg, S Lyr, R CMi:} For each of these stars, visual 
observations from AAVSO, VSOLJ and AFOEV databases show them to be
Miras and confirm the periods listed in the GCVS and in
\citet{MF2000}.  The data go back 80--90 years, over which time the
periods are all stable.

\end{document}